\documentclass[twocolumn,aps,prb,superscriptaddress,longbibliography]{revtex4-2}
\usepackage{amsmath}
\usepackage{amssymb}
\usepackage{graphicx}
\usepackage{color}
\usepackage[dvipsnames]{xcolor}
\usepackage{tabularx}
\usepackage{array}
\usepackage{float}
\usepackage{array}
\usepackage{multirow}
\usepackage{booktabs}
\usepackage{lipsum}
\usepackage[version=4]{mhchem}
\usepackage{threeparttable}
\usepackage{tabularx}
\usepackage{multirow}
\usepackage{hyperref}

\hypersetup{
    colorlinks=true,
    linkcolor=Blue,
    urlcolor=Blue,
    citecolor=Blue,
    pdftitle={Coexisting Paramagnetic Spins and Long-Range Magnetic Order in Ba4(Ru0.92Ir0.08)3O10},
    pdfauthor={Farhan Islam, Jiasen Guo, Zachary Morgan, Gang Cao, Feng Ye}
}

\begin{document}

\author{Farhan Islam}
\affiliation{Neutron Scattering Division, Oak Ridge National Laboratory, Oak Ridge, Tennessee 37831, USA}

\author{Jiasen Guo}
\affiliation{Neutron Scattering Division, Oak Ridge National Laboratory, Oak Ridge, Tennessee 37831, USA} 

\author{Wei Tian}
\affiliation{Neutron Scattering Division, Oak Ridge National Laboratory, Oak Ridge, Tennessee 37831, USA}

\author{Bing Li}
\affiliation{Neutron Scattering Division, Oak Ridge National Laboratory, Oak Ridge, Tennessee 37831, USA}

\author{Xudong Huai}
\affiliation{Department of Chemistry, Clemson University, Clemson, South Carolina 29634, USA}

\author{Thao T. Tran}
\affiliation{Department of Chemistry, Clemson University, Clemson, South Carolina 29634, USA}

\author{Gang Cao}
\affiliation{Department of Physics, University of Colorado at Boulder, Boulder, Colorado 80309, USA}

\author{Zachary Morgan}
\affiliation{Neutron Scattering Division, Oak Ridge National Laboratory, Oak Ridge, Tennessee 37831, USA}

\author{Feng Ye}
\email[Contact author: ]{yef1@ornl.gov}
\affiliation{Neutron Scattering Division, Oak Ridge National Laboratory, Oak Ridge, Tennessee 37831, USA}

\title{\texorpdfstring
  {Coexisting Paramagnetic Spins and Long-Range Magnetic Order \\
  in Ba$_4$(Ru$_{0.92}$Ir$_{0.08}$)$_3$O$_{10}$}
  {Coexisting Paramagnetic Spins and Long-Range Magnetic Order in Ba4(Ru0.92Ir0.08)3O10}
}

\date{\today}

\begin{abstract}
We investigate the effect of dilute Ir substitution on the magnetism of the
trimer-based ruthenate Ba$_4$Ru$_3$O$_{10}$ using neutron diffraction, magnetic
susceptibility measurements, atomistic simulations, and first-principles
calculations. Neutron diffraction shows that Ir doping preserves the 
zigzag antiferromagnetic structure and the ordered-moment magnitude
of the parent compound, in which
the moments reside exclusively on the two outer Ru(2) sites of each 
$\rm Ru_3O_{12}$ trimer, while the central Ru(1) site remains nonmagnetic. The N\'eel 
temperature is reduced from $\approx\!105$~K to 84.0(1)~K upon 8\% Ir substitution, 
while magnetic susceptibility reveals a pronounced low-temperature Curie-like upturn,
indicating the coexistence of paramagnetic spins with
long-range antiferromagnetic order. Density-functional calculations shows that
Ir preferentially occupies the central Ru(1) site, where its extended $5d$ orbitals
disrupt the Ru--Ru molecular-orbital network and intra/inter-trimer exchange
pathways. Atomistic simulations incorporating this paramagnetic dilution
reproduce the suppressed ordering temperature and the coexistence of ordered
and paramagnetic components. 
\end{abstract}
\maketitle

\section{Introduction}

The structural diversity of ruthenium-based oxides has made them a central
platform for exploring emergent electronic and magnetic behavior in correlated
4$d$ materials~\cite{Dussarrat96,Donohue65,Gebreyesus22,Gupta20,Nguyen19,Maeno94,Mackenzie03}.
Depending on the RuO$_6$ octahedra connectivity, ruthenates span a wide range of
physical regimes: layered compounds with corner-sharing octahedra such as Sr$_2$RuO$_4$ 
support unconventional superconductivity~\cite{Maeno94, Mackenzie03}, while other
members of the family incorporate face-sharing or edge-sharing octahedra
that assemble into dimers, trimers, or larger cluster
units~\cite{Bergemann03,Rice95,Hicks14,Yonezawa13,Kallin12}. 
These cluster-based materials are particularly intriguing because the shortened
metal--metal distances within face-sharing units promote Ru--Ru overlap,
enabling the formation of molecular-orbital--like electronic states that compete
with conventional crystal-field--split ionic configurations. This internal
electronic differentiation can modify the magnetic exchange pathways,
producing anisotropic interactions, quasi-one-dimensional behavior, or unusual
temperature-dependent crossovers between localized and itinerant  
regimes~\cite{Klein11,Igarashi13,Sannigrahi21,Hayashida25,Streltsov18,Ali24}.

\begin{figure}[!hb]
\centering
\includegraphics[width=0.9\linewidth]{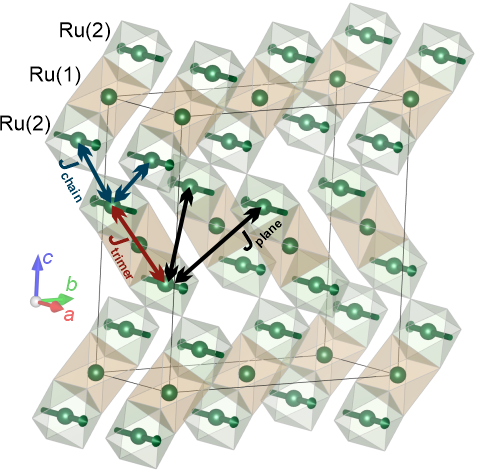}
\caption{
Magnetic structure of parent and Ir-doped Ba$_4$Ru$_3$O$_{10}$, showing the
three-dimensional network of face-sharing RuO$_6$ octahedra; oxygen and barium
atoms omitted for clarity. The face-sharing geometry forms Ru$_3$O$_{12}$
trimers composed of a central Ru(1) ion and two outer Ru(2) ions. 
The arrows illustrate the dominant magnetic exchange pathways: the intra-trimer
coupling $J_{\rm trimer}$ (red), the inter-trimer chain coupling $J_{\rm chain}$
(blue), and the in-plane inter-trimer coupling $J_{\rm plane}$ (black). 
}
\label{Fig:Structure}
\end{figure}

In face-sharing ruthenates, the magnetic behavior is influenced by the cardinality
and internal structure of the Ru--O cluster units. In dimer-based systems,
intersite $d$–$d$ hopping within
face-shared RuO$_6$ pairs can promote the formation of covalent
molecular orbitals, partially excluding electrons from the magnetic
subsystem and suppressing the effective moment~\cite{Streltsov16}.
When local moments persist on the dimers, the magnetic ground state
is controlled by the competition between strong intradimer exchange
and inter-dimer coupling, with long-range order emerging when the
two energy scales become comparable~\cite{Miiller11,Terasaki17,
Senn13,Rijssenbeek99,Jin08,Doi02}.
Trimer-based face-sharing ruthenates introduce additional internal degrees of
freedom associated with three coupled Ru ions, leading to a richer internal organization of electronic
and magnetic character among the Ru sites within the cluster~\cite{Streltsov12,Basu20}.
In addition to this internal trimer physics, it is naturally expected that the
connectivity between Ru$_3$O$_{12}$ trimers also plays an important role in determining
the magnetic behavior, as it sets the pathways through which magnetic interactions
propagate between clusters.
For example, in lanthanide-containing trimer ruthenates, linkage through
intervening $Ln$O$_6$ octahedra modifies the magnetic response~\cite{Shimoda10}.
More generally, these observations emphasize that both the delocalized electronic
configuration of Ru$_3$O$_{12}$ trimers and their inter-trimer connectivity need
to be considered to understand the complex magnetic responses in this class of materials.

Ba$_4$Ru$_3$O$_{10}$ (BRO) provides a particularly well-defined platform for
investigating magnetism in cluster-based $4d$ systems. 
BRO adopts a simple orthorhombic structure composed of face-sharing RuO$_6$
octahedra forming Ru$_3$O$_{12}$ trimers, which are further connected through
corner-sharing pathways and support three-dimensional antiferromagnetic long-range
order [Fig.~1]. Magnetically, BRO is intrinsically unconventional: only the
outer Ru(2) sites develop ordered moments, while the central Ru(1) site remains
nonmagnetic, revealing site-selective magnetism associated with the trimer-based
electronic structure~\cite{Klein11,Igarashi13,Sannigrahi21}.

The suppression of magnetic moment at the Ru(1) site was rationalized through
different theoretical proposals. First, it was proposed to originate from the
strong $\rm Ru(1)O_6$ octahedral distortion, which lifts the
$t_{2g}$ degeneracy and produces a large internal splitting ($\sim\!0.4$~eV).
The four $d$-electrons occupy the two lowest $t_{2g}$ levels, leading to a
nonmagnetic configuration at Ru(1). In contrast, the Ru(2) octahedron is less
distorted, producing only weak $t_{2g}$ splitting ($\sim\!0.15$~eV), allowing
Hund's exchange to stabilize an $S = 1$ state with a $\sim$2 $\mu_B$ moment
\cite{Sannigrahi21}.
Alternatively, the combination of molecular–orbital (MO)
framework and charge redistribution developed in
Refs.~\cite{Streltsov12,Igarashi13} provides a explanation for the
site–selective magnetism in BRO. In a face–sharing Ru$_3$O$_{12}$ trimer, the
three Ru $a_{1g}$ orbitals form bonding (B), nonbonding (NB), and antibonding
(AB) molecular orbitals. The central Ru(1) ion only participates in the B and NB
molecular orbital and the corresponding $a_{1g}$ states are pushed away from the
Fermi level, while the NB molecular orbital involves the two outermost Ru(2)
sites and forms a well defined density of states peak at the Fermi energy. In
addition, Ru(1) has more $4d$ electrons due the longer Ru--O bond lengths
\cite{Klein11}. Such charge disproportionation is confirmed by the General
Gradient Approximation (GGA) calculation \cite{Streltsov12}. This creates weaker
Ru--O hybridization and further lowers the on-site $4d$ energy (remove the Ru(1)
weight from the Fermi level). As the combined result, the lack of density of
states $N(E_{\rm F})$ for Ru(1) renders the site nonmagnetic due to the Stoner
criteria $IN(E_{\rm F}) \sim 0$, where $I$ is the Stoner electron-electron
exchange parameters~\cite{stoner47}. 

The presence of long-range AFM order in BRO despite a
magnetically silent Ru(1) site indicates that magnetic order is sustained by a
finely tuned network of exchange interactions. This is reflected in the critical
behavior: the order-parameter exponent $\beta$ is close to that of a
three-dimensional Heisenberg system, while the spin dynamics remain strongly
anisotropic, as evidenced by a sizable spin-wave gap~\cite{Sannigrahi21}. Such
features suggest that the magnetic ground state is susceptible to perturbations
that modify local exchange pathways. 
Our neutron diffraction measurements show that the Ir-substituted system
retains the same commensurate propagation vector, $\mathbf{k}=(0,0,0)$, and
zigzag AFM structure as the parent compound, while the N\'eel temperature is
suppressed to $T_{\rm N}=84(1)$~K. As in the parent material, the ordered moments
reside exclusively on the outer Ru(2) sites, whereas magnetic susceptibility
reveals a low-temperature Curie-like contribution indicative of
paramagnetic moments introduced by Ir substitution. Monte Carlo and
Landau--Lifshitz--Gilbert simulations support this interpretation, showing that
Ir dilution locally disrupts exchange interactions and suppresses $T_{\rm N}$
without altering the underlying trimer-based magnetic topology.

\section{Experimental Details}
Single crystals of Ba$_4$(Ru$_{1-x}$Ir$_x$)$_3$O$_{10}$ were grown using a
high-temperature flux method following the procedure described in
Ref.~\cite{Cao20}. Off-stoichiometric mixtures of RuO$_2$, BaCO$_3$, and
BaCl$_2$ flux were combined. For the Ir-substituted samples, IrO$_2$ was
introduced in place of a corresponding fraction of RuO$_2$ to achieve the
desired nominal doping level. The mixed powders were thoroughly ground, loaded
into platinum crucibles, and heated to ensure complete melting. The melts were
then slowly cooled to room temperature to promote single-crystal formation.
Energy–dispersive X-ray (EDX) spectroscopy was used to verify the chemical
composition of the crystals, confirming an Ir concentration of
Ba$_4$(Ru$_{0.92}$Ir$_{0.08}$)$_3$O$_{10}$ within experimental uncertainty. 

The structure information was investigated at ORNL using a Rigaku Synergy-DW
diffractometer equipped with a HyPix-Arc 150$^\circ$ detector. A rotating
molybdenum anode was used to generate X-rays with wavelength ${\lambda =
0.7107~\text{\AA}}$. The sample temperature was controlled in the range of
90-300 K by a nitrogen gas flow provided by an Oxford cryosystem. The data were
collected and reduced using Rigaku CryAlisPro software. Single crystal with
dimension of $\rm 0.05 \times 0.05 \times 0.02 mm^3$  was mounted using
Paratone-N oil on MiTeGen micro-mount. The crystal was positioned at 50~mm from
the detector. Magnetic susceptibility was measured using a Quantum Design
Magnetometer (7 T MPMS). Neutron diffraction measurements were performed using
the single crystal diffuse scattering spectrometer CORELLI~\cite{Ye18} at the
Spallation Neutron Source and VERITAS triple axis spectrometer at High Flux
Isotope Reactor of Oak Ridge National Laboratory. A single crystal with
dimension approximately $1.5\times1.5\times~0.5$~mm$^3$ was used. The sample was
glued on an aluminum pin and cooled using a closed-cycle refrigerator, with
temperature controlled from 5 to 300~K. the sample was oriented with $[0,0,L]$
out-of-plane aligned with the primary rotation axis. Three-dimensional
reciprocal-space volume data were collected at CORELLI at both 5~K and 120~K by
rotating the crystal 360$^\circ$ at 1.5$^\circ$ step. For VERITAS, reflections
accessible in the scattering plane are collected for the magnetic structure
refinements. 

\section{Results and Discussion}

The $T$-dependence of the magnetic susceptibility of parent and Ir-doped BRO
are compared in Figs.~2(a)-2(b). For the parent compound, the magnetic
susceptibility shows strong anisotropy along three crystallographic axes.
$\chi_a(T)$ displays the most pronounced kink at 105 K and subsequent suppression at
lower temperature, consistent with a spin easy-axis along the $a$ direction. In the
Ir-doped sample, the kink anomaly along $a$ shifts to a lower temperature at 84~K. In
addition, a pronounced Curie-like upturn develop for all field orientations at low
temperature.

\begin{figure}
\centering
\includegraphics[width=0.9\linewidth]{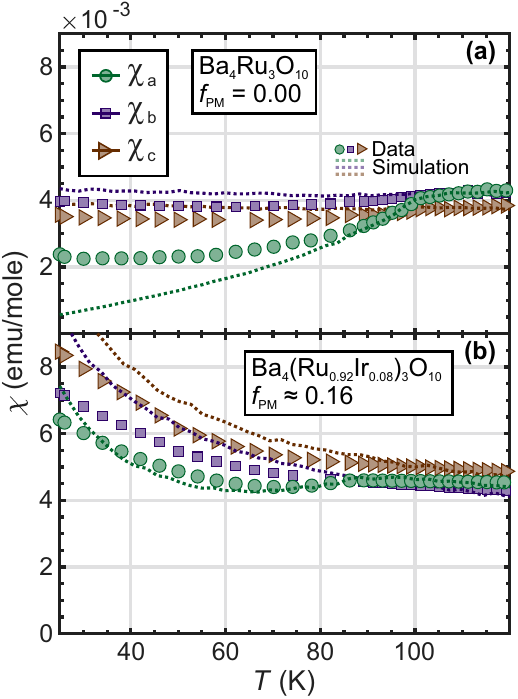}
\caption{
Temperature-dependent magnetic susceptibility $\chi(T)$ of
(a) Ba$_4$Ru$_3$O$_{10}$ and
(b) Ba$_4$(Ru$_{0.92}$Ir$_{0.08}$)$_3$O$_{10}$,
measured with magnetic fields applied along the three crystallographic axes.
The dotted curves are results of Monte Carlo spin-dynamics simulations
described in Appendix~\ref{Appendix:Sunny}.
In the parent compound, $\chi_a(T)$ shows a pronounced anomaly at
$T_{\mathrm N}\!\approx\!105$~K and suppression at lower temperatures,
consistent with an easy axis along $a$.
Upon Ir substitution, the anomaly shifts to
$T_{\mathrm N}\!\approx\!84$~K and a Curie-like upturn develops at low
temperature for all field directions, indicating the emergence of weakly
coupled paramagnetic spins.
In the simulations, Ir substitution is modeled by diluting the exchange
network through the removal of exchange bonds associated with a fraction of
paramagnetic Ru sites.
}
\label{Fig:Susceptibility}
\end{figure}

\begin{figure}
\centering
\includegraphics[width=\linewidth]{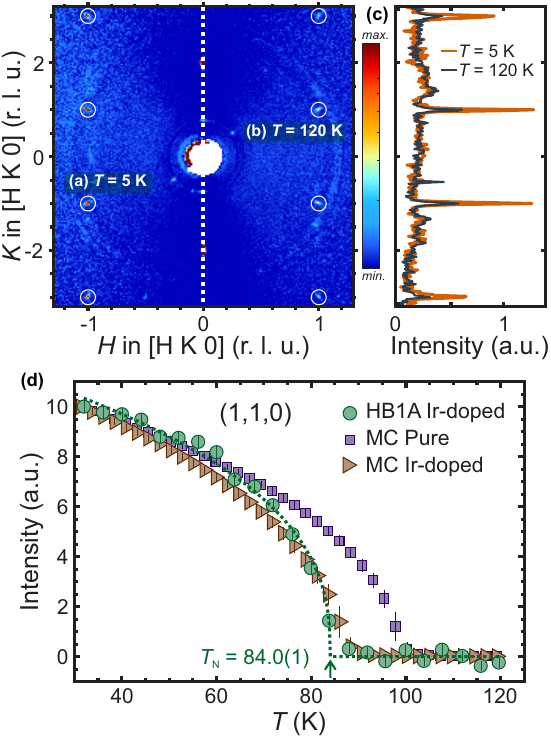}
\caption{
Neutron diffraction map of Ir-doped Ba$_4$Ru$_3$O$_{10}$ in the $(H,K,0)$ plane
at (a) $T=5$~K and (b) $T=120$~K, showing enhanced magnetic Bragg peak intensity
(circled) at low temperature. (c) Line cuts along the $K$ direction, obtained by
integrating over $H\in[0.9,1.1]$ and $L\in[-0.1,0.1]$, comparing data at $T=5$~K
and $120$~K and highlighting the enhanced magnetic intensity at the $(1,1,0)$
and $(1,2,0)$ reflections at low temperature. (d) Thermal evolution of the
integrated magnetic intensity of the $(1,1,0)$ peak measured on VERITAS (HB-1A),
fitted to extract N\'eel temperature $T_{\mathrm N}=84.0(1)$~K. Monte Carlo (MC)
spin-dynamics simulations reproduce the ordering temperatures of both the parent
compound ($T_{\mathrm N}\!\approx\!105$~K) and the Ir-doped sample. In the
simulations, Ir substitution is modeled by diluting the exchange network through
the introduction of weakly interacting paramagnetic Ru sites.
}
\label{Fig:CORELLI_SF_OP_peakCompare}
\end{figure}

To understand the evolution of the magnetic structure upon Ir-doping, 
an extensive survey in reciprocal space map was conducted at 5 and 120~K for
the Ir-doped BRO [Figs.~3(a)-3(b)]. There is no apparent structural phase
transition upon cooling. Bragg reflections can be well indexed using the reported
orthorhombic space group \( Cmca \) (No.~64) with $a = 5.67(1)$~\AA, $b =
13.245(2)$~\AA\, and $c = 13.064(2)$~\AA\ at 5~K~\cite{Klein11}. No additional super-lattice
reflection are observed. However, the peaks at low momentum transfer such as
$(1,1,0)$, $(1,3,0)$ and symmetry equivalent locations show clear enhancement in
intensity at low-$T$, which is confirmed by the line-cut comparison along the
$[1,K,0]$ direction [Fig.~\ref{Fig:CORELLI_SF_OP_peakCompare}(c)]. The resolution-limited peak width establishes
the formation of long-range magnetic order with a propagation vector $\mathbf{k}
= (0,0,0)$. The temperature dependence of the integrated (1,1,0) peak intensity
measured on VERITAS (HB1A) shows the onset of magnetic order at 84(1)~K [Fig.~\ref{Fig:CORELLI_SF_OP_peakCompare}(d)], matches well 
with the magnetic susceptibility data.

The commensurate wavevector indicates that the magnetic periodicity is the same
as the crystallographic unit cell. Symmetry analysis using
\texttt{MAXMAGN}~\cite{Perez-Mato15, Aroyo06a,Aroyo06,Aroyo11} identifies eight
magnetic space groups (MSGs) that are compatible with the crystal structure.
Details of the magnetic structure refinement procedure are provided in
Appendix~\ref{Appendix:MagneticStructureRefinement}. Among all models tested,
MSG \textit{Cmc\('\)a}, depicted in Fig.~\ref{Fig:Structure}, provides the best
refinement with goodness of fit $=0.89$. The magnetic structure preserves the
same zigzag AFM configuration as in the parent compound~\cite{Sannigrahi21},
with the ordered moment for Ru(2) is 1.01(6) $\mu_{\rm B}$, and nonmagnetic
central Ru(1). It is worth noting that the refined magnetic structure places the
ordered moments along the crystallographic $a$ axis, consistent with previous neutron
diffraction results~\cite{Klein11}, but contrasts with the $c$-axis moment
direction proposed earlier~\cite{Klein11}, which lacked access to in-plane
magnetic Bragg reflections such as $(1,1,0)$.

While neutron diffraction shows that Ir substitution preserves the zigzag
AFM ground state, moment size and orientation of the parent compound,
bulk magnetic measurements reveal quantitative differences. In particular, the
N\'eel temperature is reduced and the susceptibility develops a Curie-like
upturn. These observations indicate that dilute Ir substitution perturbs the
magnetic exchange network without altering the overall magnetic topology. To
uncover the microscopic origin of the suppressed $T_{\mathrm N}$ and the
additional paramagnetic response, we combine electronic-structure calculations
with atomistic simulations.

Spin-polarized density-functional calculations is performed for both the parent
and Ir-doped Ba$_4$Ru$_3$O$_{10}$ (Appendix~\ref{Appendix:DFT}). In the doped
case, one of the twelve Ru sites was replaced by Ir, corresponding to an
effective substitution level of 8.3\%. Three configurations were considered: the
undoped structure, Ir substitution at the central Ru(1) site, and Ir
substitution at an edge Ru(2) site. Total-energy calculations
(Table~\ref{tab:DFT_energies}) indicate that substitution at the Ru(1) site is
energetically favored among the doped cases. Atomic force and
stress analysis further reveal enhanced local distortions near the dopant site,
consistent with the more spatially extended nature of Ir $5d$ orbitals. Figures~4(a)-4(b) show the spin-density
iso-surface plots of the parent and doped compounds. A large contrast between
the middle and edge atoms is evident, underscoring the magnetic distinction
between the two Ru sites. The Ru(2) sites exhibit much large spin imbalance
comparing to the Ru(1) one, confirming their role as magnetic hosts. The spin
density shows no clear asymmetry between the two Ru(2) edge sites, which is
consistent with Ir primarily occupying Ru(1). Furthermore, the spin imbalance is
negligible at the Ru(1) site in the doped case, revealing that Ir remains
magnetically silent. This behavior is consistent with reduced moment in the 5$d$
iridates due to enhanced spin-orbit coupling and extended 5$d$
orbitals~\cite{Matsuura13}. 

\begin{figure}[b]
    \centering
    \includegraphics[width=\linewidth]{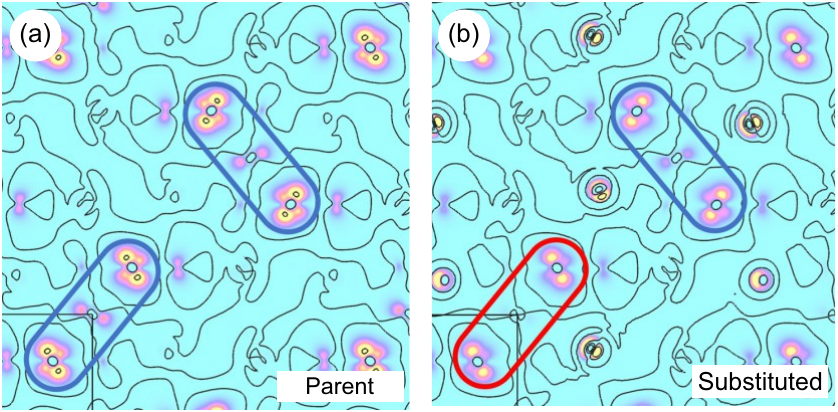}
    \caption{
        Spin-density maps for (a) the parent and (b) the Ir-substituted
        Ba$_4$Ru$_3$O$_{10}$ compound. Blue and red elliptical outlines mark the
        parent and Ir-substituted Ru$_3$O$_{12}$ trimers, respectively. In the
        parent system, the spin density is concentrated on the two outer Ru(2)
        sites of each trimer, while the central Ru(1) site exhibits negligible
        spin polarization, highlighting the site-selective nature of magnetism.
        The substituted Ru(1) site shows negligible spin polarization,
        indicating that Ir remains magnetically silent.
    }
    \label{fig:spin_density_map}
\end{figure}

Based on the preferred Ir substitution sites inferred from the electronic
structure calculations, we employed Monte Carlo (MC) and
Landau--Lifshitz--Gilbert (LLG) spin-dynamics approaches to determine the magnetic
ground state and finite-temperature magnetic response. The LLG calculations were
performed using the \textsc{Sunny} package~\cite{Dahlbom25}. Ir substitution 
occur preferentially at the central Ru(1) site, where it locally disrupts the
intratrimer exchange interaction \( J_{\mathrm{trimer}} \) that couples the two
outer Ru(2) ions within a Ru$_3$O$_{12}$ trimer. The intertrimer exchange
interactions \( J_{\mathrm{chain}} \) and \( J_{\mathrm{plane}} \), which
connect neighboring trimers along the zigzag chains and within the
crystallographic plane, are likewise treated as locally disrupted for the same
trimer. As a result, the Ru(2) ions adjacent to an Ir-centered trimer behave as
paramagnetic \( S=1 \) moments. Ir doping was implemented by introducing a
fraction \( f_{\mathrm{PM}} \) of paramagnetic Ru(2) sites, for which all
exchange interactions with neighboring magnetic ions were removed. For 8\% Ir
substitution, this dilution mechanism corresponds to a paramagnetic
Ru-spin concentration of \( f_{\mathrm{PM}} \approx 0.16 \).
Details of the spin Hamiltonian, simulation protocols, exchange parameters, and
the implementation of Ir-induced dilution are provided in Appendix~\ref{Appendix:Sunny}.

Using the exchange parameters reported in Ref.~\cite{Sannigrahi21}, 
the simulations reproduce both the zigzag antiferromagnetic ground state and the
temperature dependence of the magnetic intensity in the parent and
Ir-substituted compounds [Fig.~\ref{Fig:CORELLI_SF_OP_peakCompare}(d)]. The
corresponding magnetic susceptibilities obtained from the same models are shown
as dotted curves in Fig.~\ref{Fig:Susceptibility}. The calculations capture the
key experimental trends for both compositions, including the suppression of the
N\'eel temperature upon Ir substitution, the reduced low-temperature magnitude
of $\chi_a(T)$, and the emergence of a Curie-like upturn associated paramagnetic
Ru spins.

Ir represents a particularly strategic substitutional
perturbation for this system. According to Shannon’s revised ionic
radii for octahedral coordination, Ru$^{4+}$ (0.62${\rm \AA}$) and
Ir$^{4+}$ (0.625${\rm \AA}$) possess
nearly identical effective ionic sizes~\cite{Shannon76}, implying minimal structural mismatch
and preserving the underlying lattice framework upon substitution.
In an earlier thermodynamic and transport study, the effects of Ir substitution
on the magnetic and electronic properties are investigated for polycrystalline
$\rm Ba_4(Ru_{1-\textit{x}}Ir_\textit{x})_3O_{10}$ ($0.00\lesssim
x\lesssim0.17$)~\cite{Igarashi15}. The magnetic susceptibility are fit to a
dimer model comprised of the interacting Ru(2)--Ru(2) spins, the authors
concluded that the main effect of Ir doping is a monotonic reduction in the
intra-dimer interaction while the inter-dimer exchange coupling remains
unchanged. However, such model only describes the thermodynamics quantity near
the transition and fails to explain the apparent upturn  in magnetic susceptibility
with increasing Ir concentration far below the transition. 

On the other hand, the local disruption of exchange interactions
due to Ir substitution can be intuitively understood within the
molecular-orbital framework. In the presence of trigonal
distortion within the face-sharing Ru$_3$O$_{12}$ trimer, the Ru $t_{2g}$
manifold splits into an $a_{1g}$ orbital oriented along the trimer axis and two
$e'_g$ orbitals lying perpendicular to it. In the parent compound, magnetic
couplings involve electronic states that are delocalized over the face-sharing
Ru$_3$O$_{12}$ unit~\cite{Streltsov12}.
Substituting Ir$^{4+}$ ($5d^5$) at the central Ru(1)$^{4+}$ ($4d^4$)
site introduces an electronically distinct ion into this
delocalized network by introducing an additional $d$ electron and
enhanced spin-orbit coupling. Therefore,
the substitution perturbs the hybridization pathways that underpin
both intra- and inter-trimer exchange interactions.
Consequently, each substituted Ir ion in the Ru(1) site decouples
the adjacent two Ru spins. Further studies could clarify whether
the effects of Ir substitution arise primarily from the enhanced
spin–orbit coupling or from the additional $d$ electron. This
question could be addressed through comparative substitutions
with ions of similar size, such as Os$^{4+}$ ($5d^4$) and Rh$^{4+}$ ($4d^5$).

Despite the local disruption of magnetic exchange,
the preservation of long-range order can be understood from two
complementary considerations. First, first-principles calculations for the
parent compound show that competing magnetic configurations are separated by
substantial energy scales~\cite{Streltsov12}, reversing the relative
alignment within a Ru$_3$O$_{12}$ trimer incurs an energy cost on the order of
tens of meV per formula unit, indicating that the zigzag AFM state is
energetically favorable against local perturbations. Second, the fraction of Ru
sites rendered paramagnetic by Ir substitution ($\sim$16\%) remains well below
the percolation threshold of the three-dimensional exchange network
to disrupt the exchange network. Thus, the dilution primarily generates paramagnetic
local moments without destabilizing the global magnetic configuration.

Both bond valence sum~\cite{Igarashi13} and first-principles
calculations~\cite{Streltsov12} imply that the $\rm Ru_3O_{12}$ trimers in BRO
exhibit significant charge disproportionation between crystallographically
inequivalent Ru sites ($\sim\!0.52$ electron per atom). This
redistribution of charge is expected to modify the local electronic potential
and magnetic exchange interactions, and it naturally suggests the presence of an
incipient charge-density-wave (CDW)–like instability associated with the trimer
framework \cite{Cao26}. In cluster-based systems with strong metal–metal hybridization,
charge disproportionation can serve as a precursor to long-range charge order
correlations that are coupled to lattice and magnetic degrees of freedom. In
BRO, this interplay may be central to the emergence of site-selective magnetism
and the sensitivity of the magnetic ground state to modest chemical
substitution. Despite these indications, direct experimental evidence for static
long-range charge order remains elusive, partially because of the small lattice
distortions,  dynamical or even unconventional charge correlations. Future
studies using charge-sensitive probes, such as resonant X-ray scattering  will
be essential to clarify the nature of the charge correlations in this system.

\section{Conclusion}
In summary, we have shown that dilute Ir substitution in
Ba$_4$Ru$_3$O$_{10}$ provides a controlled route to perturb the trimer-based
magnetic network without altering the underlying zigzag antiferromagnetic
structure. Magnetic susceptibility measurements reveal a reduction
of the N\'eel temperature and the emergence of a Curie-like upturn at low
temperatures, indicating the presence of paramagnetic moments
introduced by Ir doping. Neutron diffraction establishes that long-range magnetic
order persists in the doped compound, with the ordered moments remaining confined
to the outer Ru(2) sites and the magnetic structure unchanged from that of the
parent compound. First-principles calculations show that Ir preferentially occupies the
central Ru(1) site of the Ru$_3$O$_{12}$ trimer. This site-selective substitution
disrupts the intratrimer interactions and reduces the magnetic connectivity of
neighboring Ru(2) ions, effectively releasing them as paramagnetic $S=1$
moments.  Monte Carlo and Landau--Lifshitz--Gilbert spin-dynamics
simulations further reproduce the key features of the magnetic
susceptibility in the Ir-doped system. This work demonstrates
how targeted substitution at electronically inactive sites can selectively tune
magnetic properties and  establish cluster-based oxides as the archetypal
platform for engineering coexisting ordered and paramagnetic degrees of freedom.

\section*{Ackowledgements}

G.C. acknowledges the support by the U.S. National Science Foundation via Grant
No. DMR 2204811. X.H. and T.T.T thank the Arnold and Mabel Beckman Foundation,
the NSFAward NSF-DMR-2338014, and the Camille Henry Dreyfus Foundation. This
research used resources at the High Flux Isotope Reactor and Spallation Neutron
Sources, DOE Office of Science User Facilities operated by the Oak Ridge
National Laboratory. The beam time was allocated to CORELLI on Proposal No.
IPTS-34609 and VERITAS on Proposal No. IPTS-33477. ORNL is managed by
UT-Battelle, LLC, under Contract No. DE-AC0500OR22725 for the U.S. Department of
Energy. The U.S. Government retains and the publisher, by accepting the article
for publication, acknowledges that the U.S. Government retains a nonexclusive,
paid-up, irrevocable, worldwide license to publish or reproduce the published
form of this manuscript, or allow others to do so, for the U.S. Government
purposes. The Department of Energy will provide public access to these results
of federally sponsored research in accordance with the DOE Public Access Plan.

\section*{Data Availability}

The data that support the findings of this article are not
publicly available upon publication because it is not techni-
cally feasible and/or the cost of preparing, depositing, and
hosting the data would be prohibitive within the terms of this
research project. The data are available from the authors upon
reasonable request

\clearpage

\appendix

\section{Magnetic Structure Refinement}
\label{Appendix:MagneticStructureRefinement}

The magnetic structure of Ir-doped Ba$_4$Ru$_3$O$_{10}$ is determined using neutron diffraction
measurements on the VERITAS triple-axis spectrometer at the High Flux Isotope
Reactor. A series of rocking scans were collected at accessible Bragg positions
both below ($T\!=\!5$~K) and above ($T\!=\!100$~K) the magnetic transition ($T_{\rm
N}=84.0(1)$ K). The resulting scans were fitted with Gaussian profiles to extract
integrated intensities with proper correction of instrument resolution in the triple-axis 
mode of the spectrometer. The magnetic scattering intensity was obtained by
subtracting the high-$T$ intensity from the corresponding low-$T$ value. 
The list of extracted magnetic peaks, as listed in Table~\ref{Tab:MagPeaks},
formed the experimental observables for the magnetic structure determination.

\begin{table}[b]
\centering
\renewcommand{\arraystretch}{1.2}
\begin{tabularx}{\linewidth}{>{\centering\arraybackslash}X
                            >{\centering\arraybackslash}X
                            >{\centering\arraybackslash}X
                            >{\hskip 10pt}c<{\hskip 10pt}
                            >{\centering\arraybackslash}X
                            >{\centering\arraybackslash}X
                            >{\centering\arraybackslash}X}
\toprule
\toprule
\( \mathbf{q} \) & \( F^2_{\text{calc}} \) & \( F^2_{\text{obs}} \) &
& \( \mathbf{q} \) & \( F^2_{\text{calc}} \) & \( F^2_{\text{obs}} \) \\
\midrule
\( (0\ 0\ \bar{2}) \)         & 19.21 & 22(5) &
& \( (1\ 3\ \bar{1}) \)          & 2.27 & 2.6(7) \\
\( (0\ 4\ 2) \)               & 6.21  & 4.4(8)  &
& \( (\bar{1}\ \bar{3}\ \bar{1}) \) & 2.27 & 1.9(5) \\
\( (0\ 4\ \bar{2}) \)         & 6.21  & 5.5(8)  &
& \( (\bar{1}\ \bar{3}\ 1) \)          & 2.27 & 0.8(7) \\
\( (0\ \bar{4}\ \bar{2}) \)   & 6.21  & 4.2(9)  &
& \( (\bar{1}\ \bar{5}\ 0) \)    & 1.72 & 0.0(0) \\
\( (\bar{1}\ 3\ 0) \)         & 2.79  & 0.0(0)  &
& \( (\bar{1}\ 1\ 0) \)          & 1.51 & 1.1(6) \\
\( (1\ 3\ 0) \)               & 2.79  & 2.5(5)  &
& \( (1\ \bar{1}\ 0) \)          & 1.51 & 0.0(0) \\
\( (\bar{1}\ \bar{3}\ 0) \)   & 2.79  & 0.0(0)  &
& \( (\bar{1}\ \bar{1}\ 0) \)    & 1.51 & 1.5(3) \\
\( (\bar{1}\ 3\ \bar{1}) \)   & 2.27  & 1.1(6)  &
& \( (\bar{1}\ \bar{7}\ 0) \)    & 0.21 & 0.0(0) \\
\( (1\ 3\ 1) \)               & 2.27  & 1.0(5)  &
& \( (0\ 4\ \bar{3}) \)          & 0.03 & 0.6(5) \\
\bottomrule
\end{tabularx}
\caption{Calculated and observed magnetic structure factors at selected Bragg
reflections, sorted in descending order of magnetic intensity.}
\label{Tab:MagPeaks}
\end{table}

The intensities at $T\!=\!100$ K were used to refine the nuclear structure and 
obtain the scale factor for the low-$T$ data where Bragg intensities contain
both magnetic and nuclear contribution. The refinement using the
\texttt{FullProf}~\cite{Rodriguez-Carvajal93} suite included the thermal
displacement parameters, scale factors, and extinction corrections. These
parameters were then fixed for all subsequent magnetic refinements to ensure
consistency across models.

\renewcommand{\arraystretch}{1.2}
\begin{table}[h]
\centering
\begin{tabularx}{\linewidth}{c c >{\centering\arraybackslash}X >{\centering\arraybackslash}X c}
\toprule
\toprule
&  & \multicolumn{2}{c}{Associated \texttt{IRrep}} &  \\
\cmidrule(lr){3-4}
\# & Magnetic Spacegroup & Ru(1) & Ru(2) & $\chi^2/{\rm GOF}$ \\
\midrule
1 & Cm$^\prime$c$^\prime$a$^\prime$ & -- & \(2\Gamma_2\) & 10.75 \\
2 & Cm$^\prime$ca$^\prime$          & \(2\Gamma_5\) & \(2\Gamma_5\) & 7.98 \\
3 & Cmc$^\prime$a$^\prime$          & \(1\Gamma_7\) & \(1\Gamma_7\) & 6.25 \\
4 & Cm$^\prime$c$^\prime$a          & \(2\Gamma_2\) & \(2\Gamma_3\) & 12.13 \\
5 & Cmca$^\prime$                   & -- & \(1\Gamma_1\) & 11.26 \\
6 & Cmc$^\prime$a                   & -- & \(1\Gamma_6\) & 0.89 \\
7 & Cm$^\prime$ca                   & -- & \(2\Gamma_8\) & 3.57 \\
8 & Cmca                            & \(1\Gamma_1\) & \(1\Gamma_4\) & 11.95 \\
\bottomrule
\end{tabularx}
\caption{Candidate magnetic space groups (MSGs) and associated irreducible
representations (\texttt{IRrep}s) for the Ru(1) and Ru(2) sites. A dash
indicates the absence of moment on that site.}
\label{Tab:MSG_IRrep}
\end{table}

For the propagation vector \( \mathbf{k} = (0, 0, 0) \), symmetry analysis using
the \texttt{MAXMAGN} tool~\cite{Perez-Mato15} on the Bilbao Crystallographic
Server~\cite{Aroyo06a,Aroyo06,Aroyo11} identifies eight symmetry-allowed
magnetic space groups (MSGs) that are compatible with the crystal structure of
Ba\(_4\)Ru\(_3\)O\(_{10}\). For each MSG, the \texttt{IRrep} utility in
\texttt{FullProf} was used to generate the irreducible representations
(\texttt{IRrep}) and symmetry-allowed magnetic basis vectors for the
crystallographically distinct Ru(1) and Ru(2) sites. The analysis yields four
\texttt{IRrep}s for Ru(1): \( \Gamma_1 \), \( \Gamma_3 \), \( \Gamma_5 \), and
\( \Gamma_7 \), and eight \texttt{IRrep}s for Ru(2): \( \Gamma_1 \), \( \Gamma_2
\), \( \Gamma_3 \), \( \Gamma_4 \), \( \Gamma_5 \), \( \Gamma_6 \), \( \Gamma_7
\), and \( \Gamma_8 \). These basis vectors were then matched to each MSG to
form symmetry-consistent magnetic configurations. Each magnetic model was
implemented in \texttt{FullProf}~\cite{Rodriguez-Carvajal93} and refined against
the experimental magnetic intensities. The magnetic form factor of available Ru$^{1+}$ 
\cite{prince06} was used. The quality of each refinement was assessed by the goodness of fit
\( \chi^2 \) which is reported in Table~\ref{Tab:MSG_IRrep}. 

\section{First-Principles Calculations}
\label{Appendix:DFT}

The electronic structure calculation was performed using the \textsc{Quantum
Espresso} (QE) package \cite{giannozzi09}.
The exchange–correlation potential was treated within the Generalized Gradient
Approximation (GGA) using the Perdew–Burke–Ernzerhof (PBE) functional
\cite{perdew08}. On-site Coulomb interactions in the Ru $4d$ manifold were
included via the GGA+U approach \cite{wang06} (U = 2.5~eV).
Projector-augmented wave (PAW) pseudo-potentials for Ba, Ru, Ir, and O were
taken from the \textsc{PSlibrary v.1.0.0} set \cite{dalcorso14}.
Self-consistent calculations employed a $6\times3\times3$ $k$-point mesh in
the irreducible Brillouin zone for both pristine and Ir-doped structures, with
plane-wave kinetic-energy cutoffs of 52.1~eV for the wavefunctions and
469.1~eV for the charge density. Spin-density maps were computed using the QE
post-processing module \textsc{pp.x}. Projected atomic charges and the density
of energy were obtained using \textsc{Lobster} \cite{maintz16,nelson20}.
Table~\ref{tab:DFT_energies} lists the energies from the DFT calculations 
which reveal that the most stable configuration with Ir-substitution is where it 
resides at the site in the middle of the trimer.

\begin{table}[htb]
\centering
\begin{tabular}{lccc}
\toprule
\textbf{Sample} & \begin{tabular}[c]{@{}c@{}}\textbf{Total Energy} \\ 
\textbf{(Ry/cell)} \end{tabular} & \begin{tabular}[c]{@{}c@{}}\textbf{Rel. Energy} \\ 
\textbf{(eV/cell)} \end{tabular} & \begin{tabular}[c]{@{}c@{}}\textbf{Form. Energy} \\ 
\textbf{(eV/cell)} \end{tabular} \\
\midrule
100 K     & -13829.32 & 3939.71  & -2.99   \\
Ir-Mid    & -14119.09 & 0.00     & -1.55  \\
Ir-Edge   & -14119.08 & 0.15   & -1.40  \\
\bottomrule
\end{tabular}
\caption{Total, relative, and formation energies from first-principles
calculations for undoped and Ir-doped Ba$_4$Ru$_3$O$_{10}$ configurations. The
Ir-Mid configuration is the most stable among the doped variants.}
\label{tab:DFT_energies}
\end{table}

\section{Spin-Dynamics and Monte Carlo Simulations}
\label{Appendix:Sunny}

We employed Monte Carlo (MC) and Landau–Lifshitz–Gilbert (LLG) spin-dynamics simulations
to determine the magnetic ground state and finite-temperature magnetic response.
The LLG calculations were performed using the \textsc{Sunny} package~\cite{Dahlbom25}.
Each Ru(2) ion is treated as an $S=1$ moment with an onsite single-ion anisotropy term and
a set of exchange interactions defined on the Ru$_3$O$_{12}$ trimer network. The
Hamiltonian ($\mathcal{H}$) is given by
 
\begin{equation}
\begin{aligned}
\mathcal{H}
&= 
\sum_{\langle ij \rangle \in \mathrm{trimer}}
    J_{\mathrm{trimer}} \, \mathbf{S}_i \cdot \mathbf{S}_j
\quad +
\sum_{\langle ij \rangle \in \mathrm{chain}}
    J_{\mathrm{chain}} \, \mathbf{S}_i \cdot \mathbf{S}_j
\\[2pt]
&\quad +
\sum_{\langle ij \rangle \in \mathrm{plane}}
    J_{\mathrm{plane}} \, \mathbf{S}_i \cdot \mathbf{S}_j
\quad -
D \sum_i (S_i^x)^2.
\end{aligned}
\label{Eq:Hamiltonian}
\end{equation}

with double counting bond pairs between atom \(i\)  and \(j\) omitted. The
positive sign of the exchange interaction \(J\) indicates an antiferromagnetic
coupling whereas for the single ion term \(D\) provides easy axis alignment
along the \(a\)-axis.

\begin{table}[ht]
\centering
\begin{threeparttable}
\begin{tabularx}{\linewidth}{l l c c c c c}
\toprule
\toprule
Reference & Method  & \tnote{a}$x$ & \tnote{b}$J_{\mathrm{trimer}}$ & \tnote{b}$J_{\mathrm{chain}}$ & \tnote{b}$J_{\mathrm{plane}}$ & \tnote{b}$D$ \\
\midrule
\multirow{ 2}{*}{This work}
   & \multirow{ 2}{*}{LLG \& MC}
   & 0\% & 11.5 & 4.5 & 0.5 & 2 \\
   & 
   & 8\% & 11.5 & 4.5 & 0.5 & 2 \\
\hline
Igarashi
   & \multirow{4}{*}{Dimer($\chi$ fit)}
   & 0\% & 14.6 & \multicolumn{2}{c}{\tnote{c}5.2}  & - \\
2015~\cite{Igarashi15}
   &  & 7\% & 14.3 & \multicolumn{2}{c}{\tnote{c}5.2} & - \\
   &  & 10\% & 14.1 & \multicolumn{2}{c}{\tnote{c}5.1} & - \\
   &  & 17\% & 13.8 & \multicolumn{2}{c}{\tnote{c}5.0} & - \\

\hline
Sannigrahi
   & INS(spin wave) & 0\% & 11.5 & 4.5 & 0.5 & 2.9 \\
2021~\cite{Sannigrahi21}
   & DFT(GGA) & 0\% & 14.9 & 3.7 & 1.1 & - \\
   & DFT(GGA+U) & 0\% & 11.4 & 1.3 & 0.48 & - \\
\hline
Radtke
   & DFT(GGA) & 0\% & 14.8 & 3.7 & 1.1 & - \\
2013~\cite{Radtke13}
   & DFT(GGA+U) & 0\% & 11.4 & 1.29 & 0.43 & - \\
\bottomrule
\end{tabularx}

\begin{tablenotes}
\footnotesize
\item[a] \(x\) denotes the Ir concentration in Ba\(_4\)(Ru\(_{1-x}\)Ir\(_x\))\(_3\)O\(_{10}\).
\item[b] All interaction parameters are given in the units of meV and are normalized such that \(\boldsymbol{S}\cdot\boldsymbol{S}=S(S+1)\).
\item[c] For the dimer model of Ref.~\cite{Igarashi15}, only the combined
inter-trimer exchange ($J_{\mathrm{chain}}+J_{\mathrm{trimer}}$) is reported.
\end{tablenotes}

\end{threeparttable}
\caption{Comparison of magnetic interaction parameters from the literature and
the effective values used in our Landau--Lifshitz--Gilbert (LLG) and Monte Carlo (MC) simulations.
The ratios of the exchange couplings in our model are chosen to be broadly
consistent with those inferred in Ref.~\cite{Igarashi15} while on the scale of
Ref.~\cite{Sannigrahi21}. The simulation parameters refer to the bond-paired
exchange interactions. The doped system is distinguished from the parent
compound by the fraction of paramagnetic sites $f_{\rm PM}$, which encodes the
effect of Ir substitution. Thus, assuming the underlying exchange interaction
strengths remain unchanged far from Ir-substituted sites. The resulting N\'eel
temperature and magnetic susceptibility arise as observables of the model, and
alternative parameter sets may equally well reproduce these quantities.
}
\label{tab:SunnyParams}
\end{table}

The three exchange terms in Eq.~\ref{Eq:Hamiltonian} correspond to the magnetic
pathways shown in Fig.~\ref{Fig:Structure}. The dominant coupling,
$J_{\mathrm{trimer}}$, acts along the face-sharing RuO$_6$ octahedra that form
the Ru$_3$O$_{12}$ trimer units. This interaction connects the three Ru(2) sites
within each trimer and is responsible for the zigzag AFM ordering. The second
term, $J_{\mathrm{chain}}$, couples neighboring trimers and generates
one-dimensional magnetic chains that propagate through the structure. The final
term, $J_{\mathrm{plane}}$, links trimers across the plane. The chain- and plane-
interactions are necessary for establishing the three-dimensional long-range order.
The non-magnetic Ru(1) sites are omitted. Ir substitution is incorporated by
diluting the exchange network. The Ru--Ru trimer bond is removed for the trimer
consisting of Ir atom in the middle site. Whenever such
a bond-interaction is deleted, both connected Ru sites are classified as
\emph{paramagnetic}. These sites retain $S=1$ moments but all exchange
couplings to them (trimer, chain, and plane bonds) are removed. This procedure
generates an inhomogeneous system in which weakly interacting paramagnetic
spins coexist with an AFM ordered backbone. Effectively, this dilution
procedure produces a fraction of paramagnetic Ru sites given by
$f_{\mathrm{PM}} \approx  0.24 \times 2/3 = 0.16$, once the presence
of the nonmagnetic Ru(1) sites is taken into account. This value follows from
the assumption that each substituted Ir atom occupies the central Ru(1) site
and releases two Ru(2) moments, so an Ir concentration of $x=0.08$ yields
$f_{\rm PM}=0.16$ free Ru spins.

The magnetic ground state is obtained by randomizing all spins, followed by a
finite-temperature Langevin relaxation of the LLG equation. 
For the MC simulations, a replica exchange procedure is employed over
the temperature range of interest \cite{morgan22}. At each temperature, the
system is equilibrated using Langevin dynamics or Monte Carlo Metropolis sweep.
These procedures both converge to a zigzag AFM configuration on the Ru(2) sites,
yielding a moment configuration consistent with the neutron-refined structure at
lowest temperature. The calculated magnetic order parameter is obtained from the
integrated static structure factor $I_{\mathrm{tot}}(T)=\sum_{\mathbf Q}
S(\mathbf Q)$ around the $(1\,1\,0)$ magnetic Bragg position.

The exchange parameters in our model are obtained starting with the
interaction strengths derived from inelastic neutron scattering
reported in Ref.~\cite{Sannigrahi21} cross validating that
the simulated ordering temperature reproduces the experimental value
$T_{\mathrm N}\approx105$~K for pure BRO. The $T$~dependence of the order
parameter is plotted in Fig.~\ref{Fig:CORELLI_SF_OP_peakCompare}(d) measured at HB1A and compared with
MC simulation. The single-ion anisotropy is lowered slightly to better
model the transition temperature while reducing the Ising-like
temperature-dependence obtained from the original parameters. It is noted that
for both the LLG and MC simulations, the parameters correspond to an effective
spin dipole Hamiltonian.
The full set of interaction parameters used in these simulations is summarized in
Table~\ref{tab:SunnyParams}. The simulation with a paramagnetic-site fraction
$f_{\mathrm{PM}}\approx 0.16$ with the same exchange interactions as the
parent-compound parameters closely tracks the experimental order-parameter curve
and yields a transition temperature of $T_{\mathrm N}\!\approx\!84~\text{K}$, in
agreement with neutron diffraction. 
In contrast, the undiluted simulation ($f_{\mathrm{PM}}=0$)
using the same exchange parameters for the parent compound
reproduces the higher transition temperature of $T_{\rm N}\approx 105~\text{K}$
reported in Ref.~\cite{Klein11}. This comparison demonstrates that the dilution
of exchange network contribute to the suppression of $T_{\mathrm N}$ in the Ir-doped system.

Directional susceptibilities $\chi_a(T)$, $\chi_b(T)$, and $\chi_c(T)$ were
calculated from the magnetization fluctuations. At each temperature, the system
was equilibrated for 100,000 Langevin steps and sampled for an additional 5,000
steps to obtain statistically converged susceptibility values. The calculated
susceptibilities are compared with experimental data in
Fig.~\ref{Fig:Susceptibility}. These results are validated using spin Heisenberg
Monte Carlo simulation with parallel tempering; sweeps of 30,000 for
equilibration and 20,000 for sampling are used, extracting the same
susceptibilities and order parameter. The LLG and MC simulations reproduce
the qualitative behavior of the susceptibility.

We have verified that the LLG spin-dynamics simulations performed using
\textsc{Sunny} yield temperature dependencies of the magnetic order parameter and
susceptibility that are qualitatively consistent with the MC results. For clarity,
only the MC data are presented.


%

\end{document}